# Dissecting the Specificity of Protein-Protein Interaction in Bacterial Two-Component Signaling: Orphans and Crosstalks

Andrea Procaccini[1,2], Bryan Lunt[3], Hendrik Szurmant[4*], Terence Hwa[3*], and Martin Weigt[1,2*]

[1]Human Genetics Foundation, Torino, Italy

[2]Center for Computational Studies and Dipartimento di Fisica, Politecnico di Torino, Torino, Italy

[3]Center for Theoretical Biological Physics and Department of Physics, University of California at San Diego, La Jolla, CA 92093-0374, USA

[4]Department of Molecular and Experimental Medicine, The Scripps Research Institute, La Jolla, CA 92037, USA

*Corresponding authors:
email: szurmant@scripps.edu, hwa@ucsd.edu, martin.weigt@hugef-torino.org

**Abstract**: Predictive understanding of the myriads of signal transduction pathways in a cell is an outstanding challenge of systems biology. Such pathways are primarily mediated by specific but transient protein-protein interactions, which are difficult to study experimentally. In this study, we dissect the specificity of protein-protein interactions governing two-component signaling (TCS) systems ubiquitously used in bacteria. Exploiting the large number of sequenced bacterial genomes and an operon structure which packages many pairs of interacting TCS proteins together, we developed a computational approach to extract a molecular interaction code capturing the preferences of a small but critical number of *directly interacting* residue pairs. This code is found to reflect physical interaction mechanisms, with the strongest signal coming from charged amino acids. It is used to predict the specificity of TCS interaction: Our results compare favorably to most available experimental results, including the prediction of 7 (out of 8 known) interaction partners of orphan signaling proteins in *Caulobacter crescentus*. Surveying among the available bacterial genomes, our results suggest 15~25% of the TCS proteins could participate in out-of-operon "crosstalks". Additionally, we predict clusters of crosstalking candidates, expanding from the anecdotally known examples in model organisms. The tools and results presented here can be used to guide experimental studies towards a system-level understanding of two-component signaling.

# Introduction

Signal transduction is carried out by myriads of protein-protein interactions. These interactions must be organized in a specific manner to convey desired signals while avoiding unintended crosstalks. Interaction specificity is particularly important in the light of amplified protein folds as those used by the cell for signaling. A prominent example in bacteria is the two-component signaling (TCS) system, which is highly amplified in the bacterial genomes [1]. Many organisms feature in excess of 100 homologous TCS systems regulating a flurry of behavioral responses to environmental and cellular conditions [2]. The core components of these systems are the sensor histidine kinases (SK), which detect input signals, and the response regulator proteins (RR), which relay the output [3]. The "message" between an SK and an RR is passed by the transfer of a phosphoryl group. Most TCS pathways are believed to involve a unique pair of SK/RR proteins [4; 5]. This is primarily achieved by correlating the interaction surfaces between the two proteins, giving rise to interaction specificity [6-8].

Many two-component systems are paired in operons with the SK and the RR found adjacently on the chromosome. Such a chromosomal organization immediately reveals the interacting protein pair, referred to as cognate pair (e.g., [9]). Also, complex bacteria commonly feature a large subset of "orphan" SK or RR proteins, for which no chromosomally adjacent partner can be identified [10]. Adding to the complexity, some pathways are branched, e.g. in *Bacillus subtilis* five distinct SKs (KinA-KinE) converge to phosphorylate a single RR (Spo0F) [11], and in chemotaxis, a single kinase CheA phosphorylates multiple RR proteins (CheY, CheV and CheB) [12]. In some instances, some levels of crosstalk have been observed even between cognate pairs (e.g., [13; 14]). Knowledge of signaling partners for orphans and crosstalks is a crucial component in elucidating a cell's wiring diagram towards a system-level understanding of cellular signaling.

We previously described a sequence-based computational approach, referred to as Direct-Coupling Analysis (DCA) [15; 16] and based on a message-passing algorithm [17]. Applying it to two-component protein databases, we demonstrated that sequence information could be exploited to extract residue-residue contacts across the interface of the interaction partners. DCA is based on the co-evolution of inter-protein contact residues. In contrast to traditional local co-variance techniques, DCA prunes the covariance values by separating out direct statistical couplings from correlations that arise indirectly through coupling chains (see Fig. S1), thereby vastly improving the accuracy of contact residue predictions. The idea of disentangling direct from indirect correlations was observed to be successful also in single protein domains [18; 19].

The same principle forms the basis of our approach to studying interaction specificity. Consider two homologous pairs of signaling proteins belonging to different pathways. Specificity can be achieved by changing the amino-acids exposed in the interaction surface without necessarily changing the tertiary structure of the proteins [6; 20]. Only the correct combination of interface residues in the two potentially interacting proteins leads to an actual complex formation, and results in possible phosphotransfer between these proteins. These possible combinations can be viewed as a *molecular interaction code* between signaling proteins. This code is captured in our approach by a statistical scoring scheme involving the magnitude of the direct inter-residue couplings, which is the core output of DCA. Given a pair of SK and RR sequences, the score indicates the propensity for the two proteins to interact. Our analysis shows the interaction code to be a position-specific variant of the physical interaction map between amino acids. Application of this scoring scheme to meta-genomic TCS demonstrates that interaction partners are faithfully identified, with performance much exceeding two simpler scores based on sequence similarity or on local covariance. Applying our approach to the available bacterial genomes, our results suggest 15~25% of the cognate SK/RR pairs to be involved in crosstalks. Applying it to TCS proteins in model organisms, numerous results are generated concerning crosstalk (in *B. subtilis, E. coli*) and orphan partners (in *C. crescentus, B. subtilis*). The results agree well with most of the available experimental data, and predict additional interaction partners that are previously unknown. These results in their entirety demonstrate the feasibility of utilizing sequence information for the prediction of interaction specificity, and represent a significant step forward towards a system-level understanding of entire signaling networks in bacterial organisms.

## Results and Discussion

### Statistical coupling captures physical interactions between contact residues

8,998 paired SK/RR sequences were extracted from 769 fully sequenced bacterial genomes (see Methods). Each of these pairs, as indicated by the adjacent genomic location in shared operons, forms a cognate interacting pair. They have been used in DCA to infer a global statistical model for interacting SK/RR sequence pairs (Methods Eq. (5)). The central parameters in DCA are pair-wise interaction matrices $e_{ij}(A,B)$ describing the *direct statistical coupling* between two multiple-sequence alignment (MSA) columns $i$ and $j$ (i.e. one position in each protein domain) via the amino acids $A$ and $B$ found at these positions, see Fig. S1. These matrices give direct measures of how favorable or unfavorable certain amino-acid combinations are in the considered column pairs, and likely reflect functional constraints during the co-evolution of TCS proteins. As can be seen from the examples in Fig. S2, these matrices are strongly site-specific.

It is interesting to see the extent to which physical interactions are captured by these coupling matrices derived purely from sequence statistics. Since physical interactions between contact residues may depend on details of the structural context, e.g., the local secondary structure or the relative site-chain orientation, we averaged DCA coupling matrices over the 10 column pairs with the strongest direct coupling $e_{ij}(A,B)$, which were all shown to be real inter-protein contact pairs [16]. In the resulting matrix $e_{av}(A,B) = \frac{1}{10} \sum_{(i,j) \text{high } DI} e_{ij}(A,B)$, we expect site-specific constraints to be averaged out, leaving behind common interaction mechanisms. The average matrix is shown in Fig. 1A. As indicated by the intensities of the color-coding, the entries of $e_{av}$ have smaller absolute values than the individual matrices (Fig. S2), but some interesting features become readily visible. In contrast to the individual matrices, the average matrix is almost symmetric, i.e. $e_{av}(A,B) \approx e_{av}(B,A)$: On average, a contact of an amino acid *A* in the SK and *B* in the RR is as favorable as *B* in the SK and *A* in the RR. Also, the largest entries in $e_{av}$ appear readily accountable by the electrostatics, as indicated by the red (blue) boxes between charges of opposite (like) signs.

To characterize the common interactions quantitatively, we computed the overlap of the matrix $e_{av}$ with electrostatic, hydrophilic and hydrophobic interactions. For this purpose, a feature vector $S(A)$ is introduced for each of these interactions:

- *Electrostatic interaction:* $S(A)$ has the value 1 for positively charged amino-acids (H,K,R), -1 for the negative ones (D,E), and zero for neutral amino-acids.
- *Hydrophilic interaction:* $S(A)$ has the value 1 for the hydrophilic amino-acids (D,E,H,K,N,Q,R,S,T), and zero otherwise.

- *Hydrophobic interaction:* $S(A)$ has the value 1 for the hydrophobic amino-acids (A,C,F,I,L,M,V,W,Y), and zero otherwise.

For each of these vectors, the overlap of the average coupling matrix with the corresponding interaction mode can be expressed as $q = \pm \sum_{A,B} e_{av}(A,B) S(A) S(B)$, with the prefactor -1 for the electrostatic interaction (where the opposite charges attract), and +1 for the other two interactions (where the like "charges" attract). The value of the overlap $q$ obtained is compared to the overlap $q^{(0)}$ in a null model defined by $10^6$ random permutations of the amino acids. The strongest signal is given by the electrostatic interaction ($q = 3.71$, $q^{(0)} = 0.16 \pm 0.44$), with a p-value of $2.8 \cdot 10^{-16}$ (Z-score 8.1). The signal for hydrophilic and hydrophobic interactions ($q = 1.35$, $q^{(0)} = -0.16 \pm 0.46$, p-value $4.7 \cdot 10^{-4}$, Z-score 3.3 and $q = 0.85$, $q^{(0)} = -0.16 \pm 0.46$, p-value $1.3 \cdot 10^{-2}$, Z-score 2.2, respectively) are not as strong. This might be due to the less defined nature of these interactions and possible dual function of many amino acids. In particular, the classification of hydrophobic and hydrophilic residues is not trivial. For example, while our classification is roughly based on the scale derived by Wimley and White [21], the classification scheme by Kyte and Dolittle [22] considers Y and W as hydrophilic amino acids.

The individual coupling matrices used in the calculation of $e_{av}(A,B)$ show individual biases. While some feature signals quite similar to the average matrix, others show strong preferences for a relatively small set of residue pairings (Fig. S2). In Fig 1B we display a list of the most favorable and unfavorable residue pairings. All but one can be explained by basic interactions. In the favorable list we find pairings where the individual residues have opposing charges, opposing size or are both polar. In the unfavorable set of pairings we see the opposite, i.e. same charge, same size or a polar paired with an apolar residue. The YY pairing among the negative combinations, and the FF and FY pairings among the positive ones all derive from positions 59/104 in the joint alignment. Looking at the exemplary HK853/RR468 structure (corresponding positions 291/20) the choice at these positions are FF and it is easily observable that YY would lead to a steric clash due to the additional hydroxyl group on the SK site. Whereas the averaged interaction matrix appears relatively unbiased by structure and reflects the general physical interaction between amino acids, the individual matrices show a position-specific preference for one or another type of physical interactions.

**Identification of interacting SK/RR pairs via a scoring function**

As described in Methods, the statistical model Eq. (5) inferred from the MSA of genomic cognate TCS is used to define a scoring function $Score(S_1,...,S_{L_{SK}},R_1,...,R_{L_{RR}})$ between any SK sequence $(S_1,...,S_{L_{SK}})$ and RR sequence $(R_1,...,R_{L_{RR}})$, cf. Eq. (9). This score measures the log-likelihood of sampling the two sequences under the statistical model Eq. (5), compared to the null model Eq. (8). To check the predictive power of this scoring function, we analyzed its performance when applied to TCS proteins obtained from a *metagenomic* data (see Text S1 and Fig. S3), which is a good test dataset because it contains sequences at varying level of diversity from the cognate pairs used in the construction of the statistical model. As is detailed in the Supporting Text S1, the scoring function has a very good predictive power, which is strongly increased compared to simpler schemes based on covariance alone (mutual information without DCA) or on sequence similarity. This includes an area of 91% under the receiver-operating curve compared to 71% by mutual information and 73% by sequence similarity; see Fig. S3.

**Crosstalk between cognate TCS**

Using the scoring function (Eq. (9) of Methods), we address the question of *interaction specificity* for TCS proteins extracted from the genomic library. Given that the genes for many SK/RR pairs (i.e., 70% of all SK and 44% of all RR) are located inside the same operon, how specific is this interaction kept to within the cognate (defined as chromosomally adjacent) partner and generally assumed to be interacting? Is there crosstalk between different cognate pairs inside one bacterium?

Comparing the histograms of scores obtained between cognate SK/RR partners with those for all non-cognate SK/RR pairings within the same genome (Fig. 2), it is evident that the bulk of the non-cognate SK/RR pairs (red curve) show scores well below the ones typically found for cognate SK/RR pairs (blue curve). However, the non-cognate histogram features a long tail, with a small fraction of sequences having positive scores comparable to the cognate ones (e.g., 27% of all SK with non-cognate RR with scores exceeding 20, and 14% with scores exceeding 30). This raises the possibility that some of these non-cognate SK/RR pairs reflect real inter-operon crosstalk. It should be noted that this high-scoring tail of the non-cognate histogram remains well below the cognate histogram in absolute number, suggesting that crosstalk is not very frequent even if high score does reflect interaction. In fact, out of a total of 168,332 non-cognate intra-species SK/RR pairs, only 4195 have scores above 20, 2107 above 30, and 231 above 60, whereas out of the 8998 cognate pairs 7423 score above 20, 4655 above 30, and 1092 above 60 (see Table S1 for a list of the highest scoring cross-talk candidates). In this scoring range, ~15-25% of all SK feature at least one potential crosstalk partner in addition to their cognate RR.

To address the crosstalk issue more concretely, we investigated the model organisms *Bacillus subtilis* and *Escherichia coli* in more detail. The scores between all SK and RR from cognate pairs of these organisms are represented in Figs. 3A and B, with red and blue squares indicating positive and negative scores, respectively. It is evident that the diagonals for both species, displaying the scores of cognate SK/RR pairs, are strongly red. In both figures, proteins have been arranged by hierarchical clustering [23] to group potentially cross-talking systems together. In *B. subtilis* (Fig. 3A), three sets of potentially cross-talking systems are observable, i.e. systems in which the score difference between paired and non-paired systems is small. These are the CitST/MalKR systems, the triad of BceSR/YvcPQ/YxdJK and the PhoPR/YycFG systems. These predictions match well with what has been observed *in vivo* for *B. subtilis*. Case in point, *in vivo* crosstalk has been observed between (a) PhoR and YycF (50% sequence identity with PhoP) [13] and between (b) BceS and YvcP (40% sequence identity with BceR) [14]. Equally significantly, we are not aware of any reports of crosstalk observed *in vivo* but not predicted by our scoring function. Together, our results on *B. subtilis* suggest that a high score between non-cognates is a good indicator of where crosstalk might occur.

In *E. coli,* we predicted fewer cases of crosstalk (Fig. 3B). The two prime candidates are the CitAB/DcuSR and the CusSR/YedVW systems. To the best of our knowledge, no reports are yet available on *in vivo* crosstalk between any of the paired systems that were part of our analysis. However, a comprehensive *in vitro* study suggested that crosstalk between the cognate *E. coli* TCS proteins are indeed rare [9]. In this study, YedV was observed to phosphorylate CusR (51% sequence identity with YedW), consistent with our data. The same study suggested that YfhA is a rather promiscuous RR subject to cross-phosphorylation by a number of kinases [9], but this is not evident from our analysis.

**Finding orphan interaction partners**

Another application of the scoring function is the identification of interaction partners for orphan SK and RR proteins, i.e., SK or RR genes, which are located in isolation in the genome and consequently do not have easily identifiable interaction partners. In many bacterial species, a considerable fraction of the sensing kinases are orphans, see e.g. Ref. [10]. A predictive approach that could elucidate their interaction partners would help to reconstruct the signaling network in these bacteria, a crucial step towards a systems level understanding of these organisms.

Orphan SK and RR sequences were extracted from the databases as described for the cognate systems, and scored against each other. As a general tendency, the different phylogenetic histories of interacting pairs inside or not inside operons lead to slightly lower score values for the orphans than those found for the cognate systems. We focused on *Caulobacter crescentus*, where several orphan SK/RR have been characterized experimentally [24; 25]. Of the total 46 identified HisKA-type SK (identified as described in Methods), 24 are fused to RR domains (so-called hybrid proteins) and these were excluded from the analysis. Of the remaining 22 SK, 8 are orphans and 14 are chromosomally paired with a RR protein. 44 RR proteins are not part of hybrid kinases. Among these, 20 were found to be orphans. Scores were calculated for all 8 orphan SK with the 20 orphan RR. In Fig. 4A each row displays the scores of a specific orphan SK with each of the 20 orphan RRs (indicated by the circles). Some of the scores (filled red and green circles) are very high. They may be expected to interact and can be compared to experimental results. Ohta and Newton [24] performed a yeast two-hybrid screen using the orphan RR DivK as the bait protein; they found physical interaction with the orphan SK PleC, DivJ, DivL and CC_1062 (CckN). For all these four SK, DivK is the highest-scoring orphan RR (filled red circles), even if the value of the score with DivL is comparably low. In [25], Skerker *et al*. exhaustively tested 3 orphan SKs (PleC, DivJ, CenK) for *in vitro* phosphotransfer with 44 RRs. PleC and DivJ were each found to interact with the orphan RRs PleD and DivK. These are in fact the two top ranking partners for PleC and DivJ according to our predictions (filled red circles). We note that

PleD has two RR domains; the score of the C-terminal one (orange dots) is much smaller, consistent with experiments that demonstrated that only the N-terminal RR domain is phosphorylated by both PleC and DivJ [26; 27]. The third orphan SK (CenK) that was tested *in vitro* has its partner RR (CenR) at rank 8. However, we note that scores of CenK with all of the orphan RRs are close to zero or negative, indicating that these pairs are not yet well-described by the correlated model Eq. (5).

Fig. 4A also shows the strongest predictions of new orphan interactions provided by the scoring function (filled green circles). SK CC_2755 is paired with RR CC_2757. The system is actually organized in an operon not detected by our extraction tools: CC_2757, CC_2756, CC_2755 are neighboring genes of equal coding sense, but the intergenic distance of 264bp between CC_2757 and CC_2756 slightly exceeds our operon-search cutoff of 200bp. The scoring function managed to put them back together. CC_1062 is predicted to interact also with PleD, and CC_0586 has very high scores with DivK and PleD. We recently learned that DivK is indeed an efficient *in-vitro* phosphotransfer partner for CC_0568 (Michael Laub, personal communication), suggesting that all these predictions might be correct. For the remaining SK orphan (CC_2884) only small scores with all orphan RR (comparable to the scores of CenK) are detected and hence no strong predictions are suggested. For all orphan SK, the 5 highest-ranking RR are listed in Table S2. The orphan RR protein CtrA (shown in blue in Fig. 4A) scores negatively with all kinases, consistent with a previous report that it is the terminal target of a phosphorelay and gets phosphorylated by an Hpt domain phosphotransferase rather than by any of the orphan SK proteins [28]. Other reports suggest that CtrA also gets phosphorylated by the atypical DivL kinase [29], but results are inconclusive on whether this is truly happening *in vivo* [30]. Our scoring function suggests otherwise. In conclusion, a high positive score and/or a large gap to the next scores appear to be a good indicator for interaction (no false positives), but the scoring function is still missing some SK/RR interactions (1 false negative out of 8 positives, including the matched cognate pair).

Similar results are obtained for another testable case, the orphan kinases in *B. subtilis*. Five orphan kinases KinA-KinE exist; all phosphorylate the orphan RR Spo0F, which is a part of the sporulation phosphorelay [11]. According to our scoring system, four out of the five kinases have Spo0F as its top-scoring interaction partner; the other (KinB) has Spo0F as the 4th highest scoring RR (Fig. 4B). In this case, all the maximal scores are relatively low, reflecting probably the lower specificity of Spo0F, which interacts with at least these 5 SK proteins and the phosphotransfer protein Spo0B.

**Conclusion and perspectives**

Understanding the specificity of protein-protein interaction mediating signal transduction is an outstanding challenge of systems biology. In bacterial two-component signaling involving the interaction of sensor kinases (SK) and response regulators (RR), the problem is partially solved since we observe that more than half of the SK/RRs are located adjacently on the chromosomes, and chromosomal adjacency is known to strongly imply interaction (e.g., [9]). The availability of a large number of such known interacting sequence pairs additionally provides a solid statistical basis for developing computational methods to deduce the rules of SK/RR interaction, so that the remaining half of the interacting TCS proteins may be understood. In this study, we developed and tested such a method, using the recently validated Direct-Coupling Analysis [16] as a starting point to deduce the set of direct couplings among residues of SK and RR.

The coupling matrices describing the strongest direct contacts (Fig. S2) can be understood in terms of position-specific variants of the known rules governing physical interactions among amino acids (Fig. 1). Using a statistical scoring function (Eq. (9)) constructed based on these matrices, we made a survey of the possible prevalence of crosstalk between non-cognate TCS proteins. Our results for TCS sequences obtained from genomic libraries indicate that interactions between cognate SK/RR pairs dominate, but 15-25% of the TCS proteins might also crosstalk with non-cognate members. Applying our approach to model organisms for which TCS signaling has been most thoroughly studied experimentally, we obtain results on crosstalk (Fig. 3) and orphan prediction (Fig. 4), which compare well with the existing experimental data. We also described a plethora of new predictions (Figs. 3 and 4, Tables S1 and S2, and relevant text). For example, for the relatively well-characterized signaling system in *B. subtilis*, it will be possible to directly study the predicted crosstalk between CitS and MalR, since CitS is activated by citrate [31] and MalR phosphorylation can be tracked utilizing a *ywkA* promoter reporter construct [32]. Similar studies can be carried out for the predicted crosstalk in *E. coli* between CusS/YedW, where CusS is known to be activated by copper ions [33].

Recently, Burger et al. proposed an elegant Bayesian approach that also addresses the problem of matching orphan SK with orphan RR [34]. Their method proceeds in two steps: (i) For each trial matching, the total correlation in the MSA is estimated using a dependency-tree based approximation. (ii) Matchings are sampled, using this correlation as a weight, and the probability of having a specific SK matched to a specific RR is recorded. A distinct advantage of Burger and van Nimwegen's method is that it can be applied when no large set of known interaction partners (e.g., the cognate SK/RR pairs) is available, where the present method is not applicable. However for TCS proteins, the scoring function proposed here has two major advantages: (a) At least for the experimentally available test cases discussed here, the results generated by our scoring function produce more accurate predictions; cf. Table I and Supp. Table 2 in [34]; (b) It provides a direct scoring function, so that a score can be calculated specifically for any individual SK and RR sequences. The use of the scoring function allows to zoom into individual contacts and to identify contributions towards or against the overall interaction.

An ultimate challenge would be to capture the effect of point mutations on the interaction affinity, *e.g.*, for predicting the outcome of mutation studies or designing specificity changing mutations. It is not *a priori* clear if this level of resolution can be extracted from ~9,000 diverged SK/RR pairs. We did apply our scoring function to the experiments by Skerker et al. [20] who changed the specificity of the *E. coli* kinase EnvZ from its native RR partner OmpR to RR RstA by a set of four subsequent substitutions. We find that qualitative effects of individual mutations are predicted quite well (Fig. S4). Increased experimental efforts are required to validate the effectiveness of our scoring scheme in capturing quantitative aspects of the specificity of protein interactions down to individual contact positions.

An open question is whether the developed approach will be applicable for specificity predictions in other protein systems. Observations with reduced datasets suggest that ~1000 sequences are necessary to reach comparable contact and specificity accuracy as obtained for the TCS interactions; see the insert of Fig. S3 and Fig. S5. The continuing growth of sequencing databases with now more than 1700 published bacterial genomes (according to GOLD: http://www.genomesonline.org) suggests that our approach should already be applicable for less amplified systems.

## Methods

### Data extraction

769 bacterial genomes were scanned using HMMER 2 with the Pfam 22.0 hidden Markov models (HMM, [35]) for the following Sensor Kinase (SK) domains: "HisKA" (PF00512), "HWE_HK" (PF07536), "HisKA_2" (PF07568), "HisKA_3" (PF07730),

"His_kinase" (PF06580), and "Hpt" (PF01627), for the Histidine Kinase related C-Terminal ATPase domain "HATPase_c" (PF02518), and for the Response Regulator domain "Response_reg" (PF00072) [36]. HMMER was run with its default behavior of calculating E-values as a function of the database size, against each bacterial chromosome and plasmid file available at ftp://ftp.ncbi.nih.gov/genomes/Bacteria/.

Sensor kinase domains were accepted with E-values through 10.0 but separated first into most appropriate sensor kinase domain (selected by best E-value) and then filtered by the requirement that they contain a C-Terminal ATPase domain, which is itself filtered at an E-value threshold of 0.01. Response_reg domains were simply filtered at an E-value cutoff of 0.01. Furthermore, we have excluded hybrid proteins featuring HisKA and RR, as well as proteins on predicted operons containing multiple HisKA or RR domains from further analysis. As a result, 12,814 SK sequences and 20,368 RR sequences were identified. These sequences were automatically aligned to the corresponding HMM, resulting in two MSAs for the two domain families. The aligned sequences have length $L_{SK}$ = 87 in the case of SK, and $L_{RR}$ = 117 for RR.

Many TCS are expressed from the same operon, i.e., one can identify a large number of interacting SK/RR pairs (referred to as *cognate pairs*) exploiting their genomic location on the DNA. Using a simple operational definition of an operon as a sequence of consecutive genes of same coding sense, and with inter-gene distances not exceeding 200 base pairs, a total $M$ = 8,998 cognate SK/RR pairs were identified. Cognate SK and RR sequences were concatenated and collected in a single joint MSA. The sequences in this MSA are denoted by

$$A^a = (A_1^a, ..., A_{L_{SK}+L_{RR}}^a), a = 1, ..., M \qquad (1).$$

Out of the other 11,370 RR exactly 2,334 RR were found in operons with other histidine kinase domains (HisKA_2, HisKA_3, HWE_HK) or histidine-containing phosphotransfer domains (HPt). The remaining 3,816 SK and 9,036 RR sequences are considered as *orphans*, i.e., as signaling proteins with isolated positions on the genome. Due to this filtering procedure, some SK or RR with cognate partners may actually be identified as orphans due to larger inter-gene distances inside a single operon, or due partner proteins which are not well aligned to the corresponding Pfam HMMs. Examples are given in the main text. One of the major problems addressed in this work is to assign interaction partners also to these orphan proteins. The relatively larger number of orphan RRs is due to the fact that only the largest class HisKA of SK domains was included. There are also orphan RR interacting with minor SK classes, or with the Hpt domain. As a consequence, practically all orphan SK are expected to have one or more interaction partners among the orphan RR, but not necessarily vice versa.

Note that this automated procedure is expected to be very precise for cognate TCS where the co-occurrence of three domains (HisKA, HATPase_c, Response_reg) ensures the high quality of the data set. For the analysis of orphans in *Caulobacter crescentus*, we have manually corrected the exclusion of proteins like the orphan RR PleD, which features two RR domains.

**Re-weighted frequency counts for columns / column pairs of the cognate SK/RR MSA**

The proposed statistical model for interacting SK/RR pairs is based on the statistical features of the joint MSA of the cognate pairs identified before, more precisely on the counts of amino-acid frequencies in single columns and column pairs of the MSA. Each of these columns corresponds to a specific residue position in the corresponding protein structure. The simplest quantity to look at is the single-column count

$$f_i(A) = \frac{1}{M_{eff} + \lambda q}\left(\lambda + \sum_{a=1}^{M} \frac{1}{m^a} \delta_{A,A_i^a}\right), \qquad (2)$$

which determines the frequency of finding amino-acid $A$ in column $i$ of the MSA. In this equation, a regularizing pseudo-count $\lambda = 1$ was introduced together with the notation $q = 21$ for the number of different amino acids (including the gap). The Kronecker symbol $\delta_{A,B}$ equals one if $A = B$, and zero else. Further more, a re-weighting of MSA rows (protein-pair sequences) was used to account for the uneven sampling due to phylogeny and human-based biases in the selection of sequenced bacteria (e.g., for some species multiple strains are fully sequenced). To do so, we define the numbers

$$m^a = \left|\left\{b \in \{1,...,M\} \mid seqid(A^a, A^b) > 80\%\right\}\right| \qquad (3)$$

of sequences $A^b = (A_1^b,...,A_{L_{SK}+L_{RR}}^b), b \in \{1,...,M\}$, which have more than 80% sequence identity with $A^a = (A_1^a,...,A_{L_{SK}+L_{RR}}^a)$, where $a$ itself is counted. In Eq. (3), *seqid* denotes the % identity of two aligned protein sequences. Note that the same re-weighting but with 100% sequence identity would already eliminate multiple counts of the same sequence (repeated MSA rows), but the generalization to 80% sequence identity scales down also the influence of phylogenetically very closely related SK/RR pairs. The results are of this study are not very sensitive to changes in the similarity threshold in the range between 70% and 90%, where the histogram of pairwise TCS sequence identities shows a minimum, cf. supplementary Fig. S2. The effective number of sequences becomes reduced to $M_{eff} = \sum_{a=1}^{M} 1/m^a$, as used in the normalization above. In analogy, the statistical properties for column pairs $(i,j)$ are captured by the count

$$f_{ij}(A,B) = \frac{1}{M_{eff} + \lambda q}\left(\frac{\lambda}{q} + \sum_{a=1}^{M}\frac{1}{m^a}\delta_{A,A_i^a}\delta_{B,A_j^a}\right) \quad (4)$$

of the joint appearance of amino-acids $A$ and $B$ in the same row of the MSA. One might be tempted to go beyond pair counts, but the number of elements of, e.g., a triplet count $f_{ijk}(A,B,C)$ would be $q^3 = 9261$, and thus comparable to the total number $M$ of cognate SK/RR pairs. A reliable estimate of joint frequencies beyond column pairs is thus currently not possible.

**Statistical model for interacting SK/RR proteins**

It would be tempting to use directly these frequency counts for scoring newly proposed SK/RR pairs as proposed in White et al. [7]. However, as discussed in [16], the information contained in $f_{ij}(A,B)$ depends both on direct and indirect couplings of MSA columns $i$ and $j$, and putting it together for many position pairs would necessarily lead to an over-counting of direct coupling effects. In [16] a method (called direct-coupling analysis / DCA) was introduced, which allows for disentangling direct from indirect couplings, and for writing a closed form for a global statistical model of the joint distribution of sequence pairs (rows in the joint MSA of the SK/RR pairs),

$$P(A_1,...,A_{L_{SK}+L_{RR}}) \propto \exp\left\{\sum_{i<j}^{L_{SK}+L_{RR}} e_{ij}(A_i,A_j) + \sum_i^{L_{SK}+L_{RR}} h_i(A_i)\right\} \qquad (5),$$

which contains in particular pair-wise column couplings $e_{ij}(A,B)$ measuring the direct coupling of two positions in the proteins for both inter- and intra-protein residue pairs. The a priori unknown parameters $e_{ij}(A,B)$ and $h_i(A)$ have to be determined coherently with the empirical frequency counts; in order to fulfill the constraint

$$\sum_{\{A_k|k\neq i,j\}} P(A_1,...,A_{L_{SK}+L_{RR}}) = f_{ij}(A_i,A_j) \qquad (6)$$

for all inter- and intra-protein pairs of residue positions $i$ and $j$. A detailed description how this aim can be obtained is presented in [16] and [15]. Due to current limitations in computational capacities, only up to 70 MSA columns, which are involved in the most correlated position pairs, have been included into this statistical model.

At this point it is practical to introduce the *direct pair distribution* $P_{ij}^{(dir)}(A,B)$ measuring the statistical properties $i$ and $j$ would have if all indirect couplings via intermediate positions would be pruned. Fig. S1 describes how $P_{ij}^{(dir)}(A,B)$ is determined. Note that it has the correct single-position statistics,

$$\sum_B P_{ij}^{(dir)}(A,B) = f_i(A), \qquad \sum_A P_{ij}^{(dir)}(A,B) = f_j(B) \qquad (7).$$

**Null model**

The statistical model $P(A_1,...,A_{L_{SK}+L_{RR}})$ is constructed to describe the statistical properties of the MSA of cognate SK/RR pairs, i.e., of pairs of protein sequences, which are known to interact. To determine if two newly given SK and RR sequences interact, the probability of the concatenated SK/RR sequences under the model has to be compared to a null model. As a null model, a suitable randomization of the cognate MSA is proposed: The entries of each column are randomly permuted, destroying thus all correlations between column pairs. The statistical properties of the single columns are, however, conserved under this randomization procedure. In consequence, the null model has to be described by the factorized distribution.

$$P^{(0)}(A_1,...,A_{L_{SK}+L_{RR}}) \propto \prod_{i=1}^{L_{SK}+L_{RR}} f_i(A_i) \qquad (8)$$

**The log-likelihood score**

These two statistical models shall be used to decide computationally if two protein sequences, one for a SK $(S_1,...,S_{L_{SK}})$, the other for a RR $(R_1,...,R_{L_{RR}})$, actually interact. To do so, one has to score the model for interacting domains against the null model, i.e., we have to consider a log-likelihood score of the type $\log[P/P^{(0)}]$. However, the full model contains a product over all MSA column pairs, i.e., also intra-protein pairs, and thus describes also the correlations inside the single protein MSAs. These correlations are also destroyed in the null model, so the naïve log-likelihood score is measuring if (a) the sequences of the two proteins are well described by the single-protein MSAs, and (b) if the inter-protein couplings indicate affinity between the proteins. Since we are interested only in the part (b), assuming that the given protein sequences lead to valued SK and RR protein folds, the sum in the score is restricted to inter-protein column pairs:

$$Score(S_1,...,S_{L_{SK}},R_1,...,R_{L_{RR}}) = \sum_{i \in SK, j \in RR} \log\left(\frac{P_{ij}^{(dir)}(S_i,R_j)}{f_i(S_i) \cdot f_j(S_j)}\right) \quad (9)$$

Once the model is extracted from the cognate pair MSA, the computation of the score is thus computationally efficient.

## Acknowledgment

We are grateful to R.A. White and J.A. Hoch, who were involved in early stages of this project, and to M.T. Laub for communicating unpublished results.

**Figure Legends**

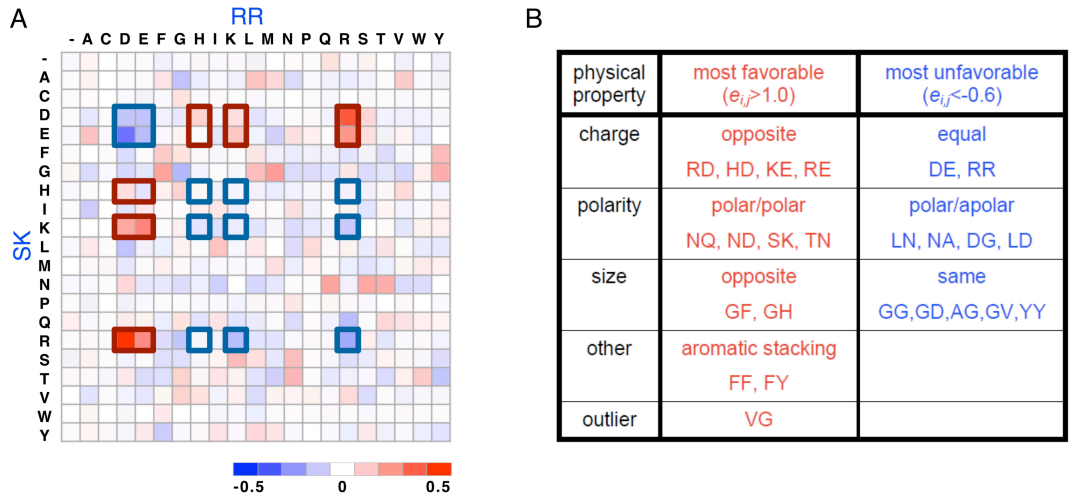

Fig. 1. Coupling matrices of the 10 strongest inter-protein couplings reveal physical interaction modes (A) Average coupling matrix $e_{ij}(A,B)$, averaged over the 10 strongest inter-protein couplings in cognate SK/RR pairs. Filled red squares indicate favorable, filled blue squares unfavorable amino-acid combinations. The matrix is shown together with the electrostatic interactions, squares inside the thick blue lines have equal charge; squares inside thick red lines have different charge and attract each other. It is obvious, that many of the strongest entries of the average interaction matrix are explainable via charge interaction. (B) List of the most favorable ($e_{i,j}$>1.0) and most unfavorable ($e_{i,j}$<-0.6) amino-acid combinations according to the 10 strongest inter-proteins couplings in cognate SK/RR pairs, together with their physical interpretation. Note that only one outlier cannot be explained in this way. For explanation of YY vs. FF see text.

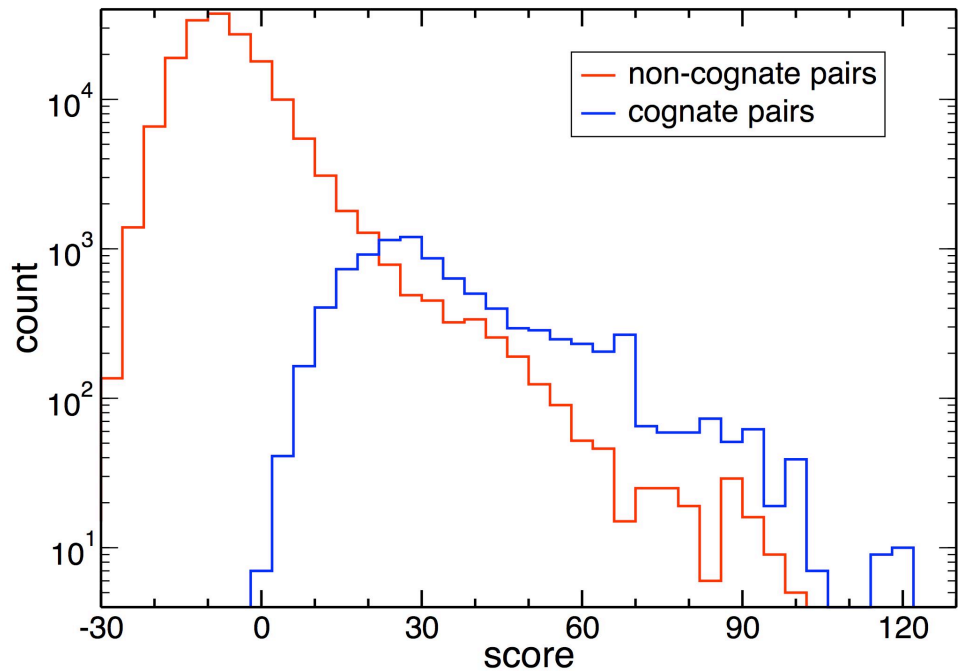

**Fig. 2. Prediction of the extent of inter-TCS crosstalk:** Histograms for intra-species pairings of all SK and all RR from cognate TCS. Shown are scores of cognate pairs (blue line) and non-cognate pairings (red line). Whereas the large majority of non-cognate pairs has scores well below the cognate ones (note the logarithmic scale), a strong tail of high-scoring non-cognate pairs unveils potential crosstalk between cognate TCS.

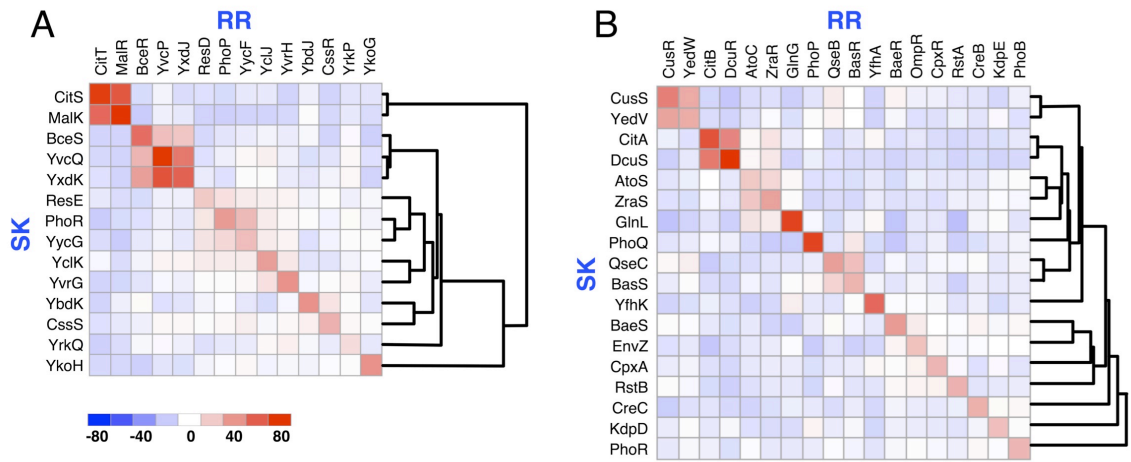

**Fig. 3. Crosstalk scores for (A)** *Bacillus subtilis* **and (B)** *Escherichia coli* **two-component systems.** Positive score correspond to red, negative to blue squares. Note that the cognate scores on the diagonal are all red as expected, but there exist some strongly red off-diagonal scores indicating potential crosstalk partners. The hierarchical clustering groups together potentially cross-talking TCS using the following dissimilarity measure: $d(TCS_1, TCS_2) = 1/\max[Score(SK_1, RR_2), Score(SK_2, RR_1)]$.

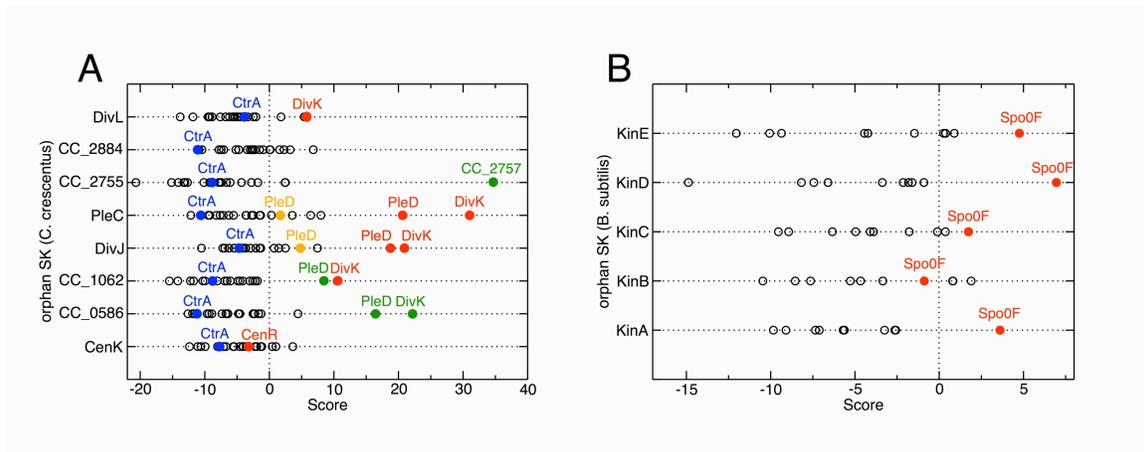

**Fig. 4. Orphan protein partner predictions (A)** Partner prediction for *Caulobacter crescentus* orphan two-component proteins. Experimentally known interaction partners [24; 25] are shown in red; the score successfully predicts 6 out of 7 interactions, including cases of known crosstalk. PleD has two RR domains, the second one has lower score (orange dots). The green dots show the best-scoring predictions of orphan interactions, the highest one (CC_2755/CC_2757) being actually a putative cognate pair, which escapes our operon search criteria. The scores of all orphan SK with the orphan RR CtrA are negative; they are represented by blue dots. **(B)** Partner prediction for *Bacillus subtilis* orphan two-component proteins. All 5 orphan kinases, KinA-E, are known to phosphorylate Spo0F [11], which is displayed in red. In 4 out of 5 cases Spo0F actually scores maximally as compared to other orphan RR (red dots). Here, we defined all RR proteins as orphans that were not in an operon with any identifiable domain from the His_kinase_A clan (CL0025 [36]). Some of the 'orphan' RR however are paired with other types of kinases or found in chemotaxis operons.

# Supporting Information Legends

**Text S1. Testing the score on metagenomic data.** The predictive performance of the scoring function is tested using cognate TCS extracted from metagenomic databases.

**Fig. S1. From pair frequency counts to direct pair distributions.** The figure shows a schematic description of the Direct Coupling Analysis (DCA). Strong correlation of the amino-acid occupation of two MSA columns *i* and *j* (detected by mutual information) may result from two different scenarios (and any mixture of the two): The two positions have a strong direct statistical coupling, or their correlation results from indirect couplings via intermediate positions. DCA disentangles direct and indirect couplings. Pruning all indirect couplings, one can determine the coupling matrices $e_{ij}(A,B)$ and the direct pair distributions $P_{ij}^{(dir)}(A,B)$, which are used in the scoring function. Technical details are explained in Weigt, M, et al. (2009) Proc Natl Acad Sci USA 106: 67-72.

**Fig. S2. Individual coupling matrices for the 9 strongest direct couplings between SK and RR positions.** The figures number all positions according to the Pfam HMMs and, in parenthesis, to the structural template HK853/RR468 (PDB ID: 3dge, Casino, P. et al. (2009) Cell 139: 325-336). Note that 25:108 was not contained in the original DCA paper (Weigt, M, et al. (2009) Proc Natl Acad Sci 106: 67-72) but showed up only in the larger cognate MSA used here. It includes the RR position 108, which, along the sequence, is far from the other identified RR interface positions. The contact 25:108 is made (3.0Å minimal atom distance) in HK853/RR468.

**Fig. S3. Prediction accuracy of the scoring function for *metagenomic* two-component system pairings.** ROC curves for the genomic training set (red) and the metagenomic test set (blue), as ordered by the interaction score. The different blue curves correspond to different subsets of the metagenomic data, which are filtered according to their dissimilarity from all genomic TCS (from top to bottom: all metagenomic TCS, metagenomic TCS featuring less than 80%, 70%, 60% resp. 50% sequence identity with all genomic TCS). The green curve is the ROC curve for the full metagenomic test set using the purely MI-based score suggested in (White, R. A. et al. (2007) Methods Enzymol 422, 75-101), the orange curve for a kNN-type scoring scheme based purely on % sequence identities ($k = 2$). The inset shows the size of the area under the metagenomic ROC curve as a function of the size of the training set (relative number of included species).

**Fig. S4. Effect of point mutations on sensor kinase/response regulator scores.** Score differences between the SK EnvZ mutants considered by (Skerker et al. (2008) Cell 133, 1043-1054) and WT EnvZ, against its native interaction partner RR OmpR and another RR RstA. The individual mutations were Mut1: L254Y; Mut2: A255R; Mut3: L254Y/A255R; Mut4: T250V/L254Y/A255R, Mut5: T250V/L254Y/A255R/S269A. Experiments by Skerker et al. demonstrated that Mut1-3 were promiscuous, phosphorylating OmpR and RstA with similar kinetics, whereas Mut4-5 preferred RR RstA over the native partner OmpR. The figure displays the score differences (= mutant score – WT score) for these substitutions. Scores of EnvZ with OmpR are found to be generally decreasing with the number of mutations (blue bars), whereas scores with RstA generally increase. This result illustrates that the scoring function is able to qualitatively capture the influence of single- and multiple- site substitutions. It is however not yet able to predict the point where specificity switches from the original cognate RR toward the new partner.

**Fig. S5. Inference of directly coupled residue pairs from reduced data sets.** Shown are the 10 strongest directly coupled residue pairs between SK and RR, for 200 (panel A), 600 (panel B), 1800 (panel C) resp. 5400 (panel D) randomly selected cognate SK/RR sequences. Residue pairs in contact are shown in red, distant pairs in green. The figure illustrates the degradation of the signal in case of insufficient sequence statistics.

**Table S1. Highest-scoring crosstalk candidate SK/RR pairs.** The table lists the name of the species, the GI numbers of the SK and the RR and the score of the two sequences, in comma-separated values (csv) file format.

**Table S2. Scores for *Caulobacter crescentus* orphans.** For each orphan SK, the 5 highest scoring orphan RRs are listed together with the scores, in csv file format.